\documentclass[aps,twocolumn,amsfonts,amssymb,pra,superscriptaddress]{revtex4-2}

\usepackage{graphicx}
\usepackage{color}
\usepackage{xcolor}
\usepackage{amsfonts}
\usepackage{amsmath}
\usepackage{braket}
\usepackage[none]{hyphenat}
\usepackage{subfigure}
\usepackage[colorlinks=true,citecolor=blue]{hyperref}
\hypersetup{colorlinks=true,citecolor=blue,linkcolor=blue,urlcolor=blue}
\usepackage{outlines}
\allowdisplaybreaks

\DeclareMathOperator{\Tr}{Tr}

\begin{document}
\title{Pathway to Kondo physics in ytterbium atom chains with repulsive spin impurities}

\author{Jeff Maki}
\email{jeffrey.maki@uni-konstanz.de}
\affiliation{Department of Physics, University of Konstanz, 78464, Konstanz, Germany}

\author{Lidia Stocker}
\affiliation{Max Planck Institute for the Physics of Complex Systems, 01187 Dresden, Germany}

\author{Oded Zilberberg}
\affiliation{Department of Physics, University of Konstanz, 78464, Konstanz, Germany}

\date{\today}

\begin{abstract}
The Kondo effect is a paradigmatic model of strongly-correlated physics, where a magnetic impurity forms a many-body singlet with a fermionic environment. Cold gases of ytterbium (Yb) atoms have been proposed to be an ideal platform to study the Kondo effect since different internal states of the atom can be used to create both the impurity and the fermionic environment. In Yb gases, however, the atomic impurity interacts with the fermionic environment both through magnetic and potential scattering. These two scattering mechanisms counteract one another, raising the question of how robust Kondo screening remains. Here, we show that potential scattering can quench the Kondo screening in one-dimensional Yb gases; yet, strikingly, Kondo physics survives this quench in well-defined regimes. Combining analytical renormalization-group theory for a Luttinger liquid with density matrix renormalization group (DMRG) simulations, we identify a transition from a strongly- to a weakly-entangled impurity as potential scattering is increased. The two approaches show excellent agreement concerning the stability of Kondo physics  throughout the different parameter regimes considered. 
Our results provide a quantitative criterion for the emergence of Kondo screening in one-dimensional Yb gases and delineate experimentally accessible regimes for its realization in cold-atom platforms.
\end{abstract}

\maketitle

\section{Introduction}

The Kondo effect arises when a localized magnetic impurity couples antiferromagnetically to a bath of fermionic quasiparticles, leading to the formation of a many-body singlet at low temperatures~\cite{Kondo1964,Anderson1970,Wilson1975}. This crossover from a weakly coupled impurity to a strongly correlated state is governed by the Kondo temperature $T_K$, a dynamically generated low-energy scale that depends nonperturbatively on the antiferromagnetic interaction strength~\cite{Anderson1970,Wilson1975}. The emergence of such a scale made the Kondo problem one of the earliest and most influential examples of renormalization-group physics in quantum many-body systems~\cite{Anderson1970,Wilson1975}.
Most famously, the Kondo effect manifests experimentally as a minimum in the resistivity of metals with magnetic impurities, followed by a logarithmic increase at lower temperatures~\cite{Kondo1964,Anderson1970,Wilson1975}. Such transport-based probes primarily reflect properties of the fermionic environment and therefore provide only indirect access to the impurity’s many-body state.

Beyond transport signatures, a variety of complementary probes have been proposed to access the many-body state of the Kondo impurity more directly, including impurity–impurity correlations~\cite{Lechtenberg2018}, charge fluctuations~\cite{Komijani2019}, entanglement measures of the impurity and its environment~\cite{Kim2021,Stocker2022,Stocker2024}, and spatial probes of the Kondo screening cloud~\cite{Stocker2025}. While these observables provide deeper insight into how the fermionic environment screens the impurity, they are often difficult to access experimentally in conventional condensed-matter settings.
Owing to its ubiquity, Kondo physics can be realized in a broad range of engineered platforms, such as quantum dots~\cite{GoldhaberGordon1998, Cronenwett1998, Pustilnik2001} and superconducting circuits~\cite{GarcaRipoll2008}. These systems offer complementary experimental control and access to impurity-specific observables, opening the door to probing aspects of the Kondo effect beyond standard transport measurements,  particularly in settings where interactions and dimensionality can be tuned with high precision.

\begin{figure}
    \centering
    \includegraphics{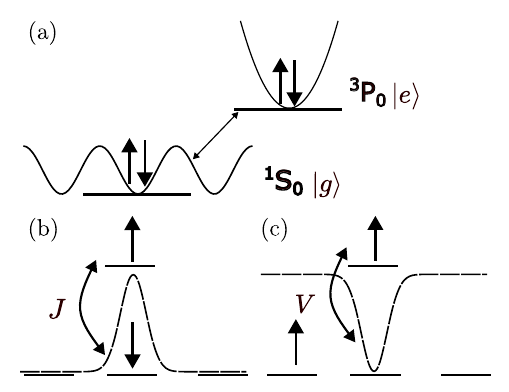}
    \caption{\textit{Kondo physics in Yb atoms.} (a) The level structure in Yb atoms exhibits a narrow clock transition between the electronic states \mbox{$^1{\bf S}_0$} and \mbox{$^3{\bf P}_0$} (marked by horizontal lines), pertaining to spin-1/2 (arrows) fermionic levels. Atoms in these states are labeled as the ground band, \mbox{$|g\rangle$}, and the excited band, \mbox{$|e\rangle$}, respectively. State selective trapping potentials (cosine vs. parabola) lead to mobile atoms in the ground band, while atoms in the excited band are localized by a strong confining potential. The resulting effective model [cf.~Eq.~\eqref{eq:Microscopic_Lattice_Model}] is an impurity in a fermionic environment of spin-1/2 fermions interacting with (b) antiferromagnetic and (c) potential interactions. These interactions tend to draw atoms (b) towards or (c) away from the impurity. The dashed lines schematically represent the modulus of the wavefunction of atoms near the impurity site when either magnetic or potential scatterings are dominant.}
    \label{fig:1}
\end{figure}

A particularly versatile platform for realizing Kondo physics is provided by cold-atom experiments with ytterbium (Yb) atoms~\cite{Gorshkov2010,Nakagawa2015,Zhang2016,KanszNagy2018,Goto2019,Wei2025,Amaricci2025}. As illustrated in Fig.~\ref{fig:1}, the atomic structure of Yb features two long-lived electronic states separated by an ultra-narrow clock transition~\cite{Boyd2007}. These states, denoted ground \mbox{$|g\rangle$} and excited \mbox{$|e\rangle$}, can be used to effectively realize two species of spin-$1/2$ fermions. By employing state-selective optical potentials~\cite{Hoehn2023}, the motion of atoms in one state can be strongly suppressed, thereby realizing localized magnetic impurities, while atoms in the other state remain mobile and form the fermionic environment. This architecture provides a highly-controllable setting for addressing central questions in Kondo physics, including impurity quench dynamics~\cite{KanszNagy2018,Goto2019,Wei2025}, which enables direct access to the quantum state of the impurity and to the entanglement between the impurity and its environment~\cite{Stocker2022,Stocker2024,Stocker2025}.

Cold-atom systems, furthermore, provide a controlled setting to explore how many-body correlations in the environment interplay with the formation of the Kondo effect. Specifically, using Feshbach resonances, the strength of interactions within the fermionic environment can be tuned over a wide range~\cite{Chin2010}. In conventional Fermi-liquid environments, repulsive correlations are known to suppress the Kondo temperature, whether they originate from interparticle interactions~\cite{Schork1994,Li1995} or from additional impurity scattering~\cite{Amaricci2025}.
In one dimension (1D), instead, interaction effects in the environment become dominant, since even weak repulsion destabilizes the Fermi liquid in favor of a Luttinger liquid (LL)~\cite{Haldane1981,Giamarchi2003,Bouchoule2025}. When a magnetic impurity is embedded in a LL, the Kondo effect is predicted to be qualitatively altered, with the Kondo temperature acquiring a nonuniversal power-law dependence on the impurity scattering strength as a consequence of enhanced back-scattering between the two Fermi points~\cite{Lee1992,Furusaki1994}.
This power-law scaling replaces the exponential dependence that is characteristic of Fermi liquids. As cold atomic gases have already established themselves as a clean and versatile platform for realizing and probing LL physics~\cite{Hofferberth2008,Fabbri2015,Yang2018,Senaratne2022,CavazosCavazos2023,Bouchoule2025}, it substantially improves the prospects for observing Kondo physics in cold-atom experiments, providing a natural motivation for the present work.

A central challenge for realizing the Kondo effect in Yb atomic gases, however,  is that the impurity couples to the fermionic environment through both magnetic exchange and non-magnetic potential scattering~\cite{Gorshkov2010,Amaricci2025}. As illustrated schematically in Fig.~\ref{fig:1}, these two interactions act in opposition and lead to a suppression of the Kondo temperature. While the antiferromagnetic exchange favors the formation of a screening cloud around the impurity, repulsive potential scattering repels atoms in the ground state from the impurity and thereby weakens Kondo correlations. As a result, potential scattering constitutes a fundamental obstacle to observing the Kondo effect in cold atomic gases~\cite{Amaricci2025}.

In this work, we resolve how Kondo screening survives in interacting one-dimensional fermionic Yb gases despite competing non-magnetic potential scattering. We demonstrate that while potential scattering can strongly suppress the Kondo temperature, distinct interaction regimes preserve robust Kondo physics. Our approach combines a Wilsonian renormalization-group (RG) analysis of a Luttinger liquid with magnetic exchange and potential scattering~\cite{Altland2010,Haldane1981,vonDelft1998,Giamarchi2003} with large-scale tensor-network and DMRG simulations~\cite{schollwock2011,Fishman2022}. Together, these methods provide direct access to the zero-temperature many-body ground state and yield a quantitative description of the Kondo screening cloud in an interacting fermionic environment.

The remainder of this article is organized as follows: In Sec.~\ref{sec:LL}, we introduce the model and its LL description in 1D. In Sec.~\ref{sec:results}, we present the main results of this work. We then provide the details of our RG analysis in Sec.~\ref{sec:RG}. Section~\ref{sec:DMRG} discusses the implementation of the DMRG method to study the ground state of the 1D Fermi-Hubbard model. We then present our conclusions in Sec.~\ref{sec:conclusions}.

\section{Luttinger liquids in ytterbium gases}
\label{sec:LL}

In this section, we introduce the effective model for realizing Kondo physics using Yb gases in a one-dimensional optical lattice~\ref{subsec:fermiHubbard}. We then proceed to write its corresponding LL description in the continuum~\ref{subsec:Luttinger}. 
\subsection{Fermi-Hubbard model with ytterbium gases}
\label{subsec:fermiHubbard}
The effective Hamiltonian for an impurity in a 1D Yb gas [cf.~Fig.~\ref{fig:1}] is the Fermi-Hubbard model with both magnetic and potential impurity scatterings~\cite{Gorshkov2010} 
\begin{align}
H &= -t \sum_{i,s} \left[c_{i+1,s}^{\dagger}c_{i,s} + h.c.\right] + \sum_{i} U n_{i,s}n_{i,\bar{s}} \nonumber \\
& +\frac{J}{2} \vec{S} \cdot \sum_{s,s'} c^{\dagger}_{0,s}\vec{\sigma}_{s,s'} c_{0,s'} +V\sum_s n_{0,s}\ ,
\label{eq:Microscopic_Lattice_Model}
\end{align}
Here, \mbox{$c_{i,s}^{(\dagger)}$} is the annihilation (creation) operator of a fermionic atom in the ground band \mbox{$|g\rangle$}, at site \mbox{$i$} and with a spin \mbox{$s=\pm 1$} (\mbox{$\bar{s} = -s$}). Additionally, \mbox{$t$} is the hopping strength between sites of a 1D optical lattice, \mbox{$n_{i,s} = c_{i,s}^{\dagger}c_{i,s}$} is the local number operator, and \mbox{$\vec{\sigma}_{s,s'}$} are the Pauli sigma matrices. The impurity is denoted as a quantum spin \mbox{$\vec{S}$}, and is assumed to preside on top of site \mbox{$i=0$} of the \mbox{$\ket{g}$}-atoms chain.

In experimental setups, four types of interactions exist; two intra-band scatterings and two inter-band scatterings. The intra-band scattering between fermionic atoms in the ground band leads to the Fermi-Hubbard interaction, \mbox{$U$}. Since we only focus on a single impurity, we can neglect similar intra-band scattering between \mbox{$|e\rangle$} atoms. The two inter-band scattering channels depend on whether the two scattering atoms have a wavefunction that is an orbital singlet or triplet (\mbox{$|e\rangle \pm |g \rangle$}). Linear combinations of these interactions lead to the magnetic impurity scattering, \mbox{$J$}, and potential terms,  \mbox{$V$}, shown in Figs.~\ref{fig:1}(b)--(c), respectively~\cite{Gorshkov2010}. 

\subsection{Continuum limit and Luttinger liquids}
\label{subsec:Luttinger}

At low temperatures and in the continuum limit, the low-energy excitations of Eq.~\eqref{eq:Microscopic_Lattice_Model} are bosonic fluctuations of either density or spin. These fluctuations can be accurately described using bosonization and the following LL model~\cite{Haldane1981,vonDelft1998,Giamarchi2003}
\begin{align}
H &= H_{LL} + \delta H_V + \delta H_{z,F} +\delta H_{z,B} + \delta H_{+} + \delta H_{-} \ , \label{eq:H}
\end{align}
where 
\begin{subequations}
\begin{align}
H_{LL} &= \sum_{\nu= \rho,\sigma} \frac{v_{\nu}}{2\pi} \int dx \left[K_{\nu} \left(\partial_x \theta_{\nu}(x)\right)^2 + K_{\nu}^{-1} \left(\partial_x \varphi_{\nu}(x)\right)^2\right],\label{eq:HLL}\\
\delta H_{V} &= V \sum_{r,s}\left[:\psi_{r,s}^{\dagger}(0) \psi_{r,s}(0) :+:\psi_{r,s}^{\dagger}(0) \psi_{\bar{r},s}(0) :\right], \label{eq:Udir} \\
\delta H_{z} &= \frac{J}{2} S^z  \sum_{r,s} s \left[:\psi_{r,s=\pm 1}^{\dagger}(0) \psi_{r,s}(0) :+:\psi_{r,s}^{\dagger}(0) \psi_{\bar{r},s}(0) :\right], \label{eq:Uex_z} \\
\delta H_{\pm} &= \frac{J}{4} S^{\mp}\sum_{r=\pm 1} \left[ :\psi_{r,\pm}^{\dagger}(0) \psi_{r,\mp}(0): + :\psi_{r,\pm}^{\dagger}(0) \psi_{\bar{r},\mp}(0):\right]. \label{eq:Uex_pm}
\end{align}
\end{subequations}
Throughout this work, we set the units \mbox{$\hbar = m = 1$}.
Eq.~\eqref{eq:HLL} is the standard LL Hamiltonian describing the linearly dispersive excitations of the ground band atoms. Due to spin charge separation, the density modes, \mbox{$\nu = \rho$}, and spin modes, \mbox{$\nu = \sigma$}, have different speeds of sound, \mbox{$v_{\nu}$}. The bosonic fields for these sound modes \mbox{$\theta_{\nu}(x)$} and \mbox{$\varphi_{\nu}(x)$} satisfy the commutation relations
\begin{equation}
    \left[\varphi_{\nu}(x),\partial_y \theta_{\nu'}(y)\right] = i \pi \delta_{\nu,\nu'}\delta(x-y) \ .
    \label{eq:commutator}
\end{equation}
The parameters \mbox{$K_{\nu}$} are known as the Luttinger parameters and depend on the Fermi wavenumber and the interactions, $U$, between atoms in the ground band. These parameters can be extracted from the underlying lattice model by calculating the static structure factor for density and spin~\cite{Hardikar2007}
\begin{align}
    K_{\rho} &= \pi \lim_{q \to 0} \frac{1}{q} \sum_{s,s'}\sum_{j,k}e^{i q(j-k)} \langle n_{j,s} n_{k,s'}\rangle \label{eq:Krho}\ , \\
    K_{\sigma} &= \pi \lim_{q \to 0} \frac{1}{q} \sum_{s,s'}s s'\sum_{j,k}e^{i q(j-k)} \langle n_{j,s} n_{k,s'}\rangle \label{eq:Ksigma}\ .
\end{align}

\begin{table}[]
\begin{tabular}{|c|c|c|}
\hline
 Interaction & Coupling Constant & Type\\ \hline
 $\delta H_{z,F}$ & $\eta $  & Magnetic\\ \hline
 $\delta H_{z,B}$& $\tilde{J}_{z,B}$ & Magnetic\\ \hline
 $\delta H_{\perp,F}$ & $\tilde{J}_{\perp,F}$ & Magnetic\\ \hline
 $\delta H_{\perp,B}$ & $\tilde{J}_{\perp,B}$ & Magnetic\\ \hline
 $\delta H_{V,B}$ & $\tilde{V}$ & Potential - charge\\ \hline
 $\delta H_{2,0}$ & $\tilde{V}_{2,0}$ & Potential - charge\\ \hline
 $\delta H_{0,2}$ & $\tilde{V}_{0,2}$ & Potential - spin \\
 \hline
\end{tabular}
\caption{Summary of the dimensionless interaction parameters appearing in the RG flows in the weakly interacting limit [cf.~Eq.~\eqref{eq:RG_equations_final}], their dimensionless coupling constants, and the type of interaction. \label{tab:1}}
\end{table}

The terms $\delta H_V, \ \delta H_z, \ \delta H_\pm$ in Eqs.~\eqref{eq:Udir}-\eqref{eq:Uex_pm}, describe the impurity scattering and are written in terms of normal ordered products of fermionic operators \mbox{$\psi_{r,s}^{(\dagger)}(x)$} that annihilate (create) either a right moving (\mbox{$r=+1$}) or left moving (\mbox{$r=-1$}) fermion with spin \mbox{$s=\pm 1$} at the position of the impurity \mbox{$x=0$}. The normal ordering of a fermionic operator is denoted as \mbox{$: \psi_{r,s}^{(\dagger)}(x) : =  \psi_{r,s}^{(\dagger)}(x) - \langle GS |\psi_{r,s}^{(\dagger)}(x)  | GS \rangle$} and requires the removal of the expectation value of an operator in the ground state, \mbox{$|GS\rangle$.}
The fermionic fields themselves are related to the bosonic fields via
\begin{equation}
    \psi_{r,s}(x) = \frac{F_{r,s}}{\sqrt{2\pi a}}e^{-\frac{i}{\sqrt{2}} \left(\left(\theta_{\rho}(x)-r \phi_{\rho}(x)\right) + s\left(\theta_{\sigma}-r \phi_{\sigma}(x)\right)\right)}\,,
    \label{eq:bosonized_fermion}
\end{equation}
where \mbox{$a$} is a short distance cutoff related to the lattice spacing in Eq.~\eqref{eq:Microscopic_Lattice_Model} and \mbox{$F_{r,s}$} is the Klein factor that keeps track of the fermionic statistics. 
Note that the impurity scattering exhibits two varieties:  forward-scattering, which conserves the number of \mbox{$r$}-movers, and backscattering, which flips an \mbox{$r$}-mover as \mbox{$r\to\bar{r}=-r$}. As we will discuss later, it is important to separate these two scattering types as they behave differently under the RG flow.
Further information about the derivation of Eqs.~\eqref{eq:H} and the process of bosonization is outlined in Appendix~\ref{app:LL}. 

As is standard in LL theory, the RG flow generates additional interaction terms that are not present in the bare Hamiltonian. This reflects the non-renormalizability of the theory, in the sense that integrating out high-energy modes produces new operators consistent with the symmetries of the problem. The leading contributions correspond to two particle–hole scattering processes
\begin{align}
    \delta H_{2,0} &= V_{2,0} \sum_{r,s}  :\psi_{r,s}^{\dagger}(0) \psi_{r,\bar{s}}^{\dagger}(0)\psi_{\bar{r},\bar{s}}(0)\psi_{\bar{r},s}(0): \,,\label{eq:deltaH20}\\
    \delta H_{0,2} &= V_{0,2} \sum_{r,s}  :\psi_{r,s}^{\dagger}(0) \psi_{\bar{r},\bar{s}}^{\dagger}(0)\psi_{r,\bar{s}}(0)\psi_{\bar{r},s}(0):  \,,\label{eq:deltaH02}
\end{align}
which we henceforth include in the Hamiltonian of the system. A complete summary of all the terms in our LL Hamiltonian is presented in Tab.~\ref{tab:1}.

\section{Results}
\label{sec:results}
In this section, we present the central results of our RG and DMRG analyses, with the methodological details and technical derivations provided in the subsequent discussion, cf. Secs.~\ref{sec:RG} and~\ref{sec:DMRG}.

\subsection{Perturbative renormalization group}
\label{sec:pertRG}
We begin by mapping out the phase diagram using a RG approach. As described in Sec.~\ref{sec:RG}, we employ the Wilsonian momentum-shell method to derive the RG flow equations in the weak-coupling regime, retaining terms up to second order in the cumulant expansion~\cite{Altland2010}. The resulting flow equations are given in Eq.~\eqref{eq:RG_equations_final}. By analyzing these equations from a fixed initial dimensionless magnetic interaction of strength \mbox{$\tilde{J}(0)=10^{-4}$} and for variable initial dimensionless potential scattering strengths \mbox{$\tilde{V}(0)$}, we identify which coupling grows most rapidly as the temperature is lowered.
When a given coupling constant reaches order unity, the perturbative RG treatment ceases to be controlled. We therefore assume that the subsequent flow is driven towards the strong-coupling fixed point associated with the first coupling to cross unity. This assumption is validated a posteriori by comparison with DMRG simulations, which directly probe the true many-body ground state for finite-size systems.

\begin{figure}
    \centering
    \includegraphics{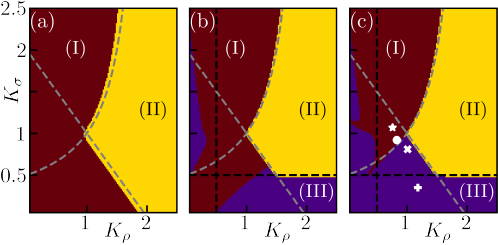}
    \caption{\textit{Phase diagram for the Kondo effect in the weakly interacting limit}. We set \mbox{$\tilde{J}(0) = 10^{-4}.$} The potential scattering is set to be: (a)--(b)\mbox{$\tilde{V}(0)/\tilde{J}(0) = 0$}  and (c) \mbox{$\tilde{V}(0)/\tilde{J}(0) = 5$}. (a) Corresponds to the RG flow of the ideal Kondo problem, (b) to the ideal Kondo problem with the leading irrelevant perturbations, and (c) the full RG flow including both potential scattering and the leading irrelevant interactions [cf.~Eqs.~\eqref{eq:RG_equations_final}]. (I) Red marks the Kondo regime,  (II) yellow the perturbative regime, and (III) purple the strong potential scattering regime dominated. The silver (black) dashed lines denote where the scaling dimensions of the magnetic interactions (potential scattering) are marginal. White $\mathbf{\cdot},{\star},{\scriptstyle \times},{\scriptstyle +}$ markers indicate the points considered in panels (i)-(iv) of Fig.~\ref{fig:5}, respectively. }
    \label{fig:2}
\end{figure}

In Fig.~\ref{fig:2}, we depict the resulting phase diagram of the magnetic impurity coupled to a LL, constructed at three successive levels of approximation. Throughout, we distinguish three types of behavior: In the Kondo regime (red), magnetic exchange interactions grow strong under the RG flow. In the strong potential-scattering regime (purple), non-magnetic impurity terms dominate. In the perturbative regime (yellow), all couplings remain weak down to low temperatures.
We begin in Fig.~\ref{fig:2}(a) with the simplest approximation, in which only magnetic impurity interactions are retained and all potential scattering processes are neglected, i.e., we fix \mbox{$\tilde{V}=\tilde{V}_{2,0}=\tilde{V}_{0,2}=0$}, cf.~Table~\ref{tab:1}. Supplementing this limit with the SU(2)-invariant condition,  \mbox{$K_{\sigma}=1$}, reduces the flow equations to the standard poor man’s scaling description of the Kondo effect in a Luttinger liquid with its critical quantum critical point at \mbox{$K_{\rho}=1$}~\cite{Furusaki1994}. The silver dashed lines indicate where the scaling dimensions of the longitudinal, \mbox{$\tilde{J}_{ z,B}$}, and transverse, \mbox{$\tilde{J}_{\perp, B}$}, exchange couplings become marginal, corresponding to the conditions \mbox{$K_{\rho}+ K_{\sigma} = 2$} and \mbox{$K_{\rho}+ K_{\sigma}^{-1}= 2$}, respectively. These lines accurately delineate the boundary between the Kondo regime and the perturbative regime in this approximation.

A consistent RG treatment to second order, however, generates additional interaction terms describing two particle-hole scattering processes, given in Eqs.~\eqref{eq:RG_V_20}-\eqref{eq:RG_V_02}; we keep $\tilde{V}(0)=0$ fixed here. The effect of including these terms is shown in Fig.~\ref{fig:2}(b); they become marginal when \mbox{$K_{\rho} =1/2$} or \mbox{$K_{\sigma}=1/2$}, as indicated by the black dashed lines. We observe that the added particle-hole scattering processes substantially reduces the parameter space in which Kondo physics survives. In particular, when \mbox{$K_{\rho}>1/2$} and \mbox{$K_{\sigma}<1/2$}, or vice versa, the RG flow is instead driven toward a phase in which either the density or phase field becomes pinned, leading to either singlet superconducting (\mbox{$V_{2,0}$}) or spin-density wave (\mbox{$V_{0,2}$}) order \cite{Furusaki1993, Giamarchi2003}. Such pinning suppresses charge or spin transport and preempts the development of Kondo screening~\cite{Furusaki1993,Kane1992b}. 

Finally, in Fig.~\ref{fig:2}(c), we include both the generated two particle-hole scattering terms and explicit potential scattering at the impurity, i.e., all the RG flow Eqs.~\eqref{eq:RG_equations_final} are used. The results are shown for a representative ratio \mbox{$\tilde{V}(0)/\tilde{J}(0) = 5$}. In the weak potential-scattering limit, \mbox{$\tilde{V}(0)/\tilde{J}(0) \ll 1$}, the phase diagram reduces to that of Fig.~\ref{fig:2}(b). Once 
\mbox{$\tilde{V}(0)/\tilde{J}(0) \sim 1$}, qualitative changes emerge and the region supporting Kondo screening is further reduced. For even larger \mbox{$\tilde{V}(0)/\tilde{J}(0)$}, the Kondo regime can even be completely suppressed and replaced by a strong-coupling phase dominated by potential scattering.

\begin{figure}
    \centering
    \includegraphics{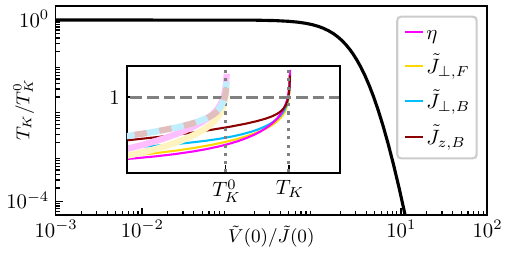}
    \caption{\textit{Kondo temperature scaling.} (a) Kondo temperature for various potential scattering strengths. The initial value of the magnetic scattering is \mbox{$\tilde{J}(0)=10^{-3}$}, and the maximum allowed initial potential scattering is \mbox{$\tilde{V}(0)=10^{-2}$}. In the inset, we plot the RG flows obtained from solving Eq.~\eqref{eq:RG_equations_final} for the magnetic interactions. The color scheme is the same as in Fig.~\ref{fig:7} below, where thin lines denote the results for \mbox{$\tilde{V}(0)/\tilde{J}(0)=100$}, while thick transparent lines are for \mbox{$\tilde{V}(0)=0$}. We use dashed alternating thick lines when results overlap and would otherwise be indistinguishable. The Kondo temperature in the presence (absence) impurity scattering is denoted by \mbox{$T_k^{(0)}$}. The Kondo temperature is defined when the magnetic interactions are equal to unity: $\tilde{J}=1$.}
    \label{fig:3}
\end{figure}

Crucially, the region characterized by \mbox{$K_{\sigma} \approx 1$} and \mbox{$1/2<K_{\rho}<1$} is found to be remarkably robust against potential scattering, i.e., it remains in the Kondo regime even for large values of \mbox{$\tilde{V}(0)/\tilde{J}(0)\sim 10$}. This robustness is nontrivial, since \mbox{$\tilde{V}$} constitutes a relevant perturbation in the renormalization-group sense and would naively be expected to destabilize Kondo screening. To elucidate this behavior, we plot the Kondo temperature as a function of \mbox{$\tilde{V}(0)/\tilde{J}(0)$} for \mbox{$K_{\rho} = 0.7$} and \mbox{$K_{\sigma}=1.0$} in Fig.~\ref{fig:3}.  We find that in this parameter regime the system always flows toward a Kondo strong-coupling fixed point, but with a Kondo temperature that is progressively reduced as \mbox{$\tilde{V}(0)$} increases. This behavior contrasts sharply with other regions of the phase diagram, highlighted in purple, where increasing \mbox{$\tilde{V}(0)$} instead drives the RG flow toward a strong potential-scattering fixed point and completely suppresses Kondo screening. Representative examples of the different RG flow trajectories are discussed in Appendix~\ref{app:RG}. Taken together, these results demonstrate that while potential scattering universally suppresses the Kondo temperature, it does not generically eliminate the Kondo effect. Instead, there exist extended interaction regimes in which Kondo screening survives, albeit at reduced energy scales. This robustness in the presence of strong potential scattering constitutes the first central result of this work.

\subsection{Ground state wavefunction from DMRG}

We benchmark our perturbative RG analysis against DMRG calculations, in the matrix product states (MPS) framework, for the many-body ground state of the microscopic Fermi-Hubbard model~\eqref{eq:Microscopic_Lattice_Model}. The DMRG approach provides an essentially exact treatment of the interacting lattice system at zero temperature and therefore offers a direct probe of the emergence of Kondo screening in the true many-body ground state.

\begin{figure}
    \centering
    \includegraphics{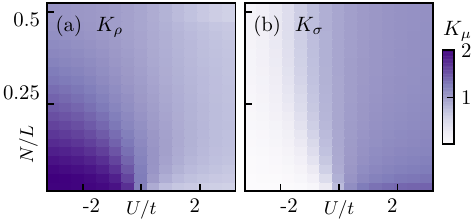}
    \caption{\textit{Luttinger liquid parameters.} We calcuate the $K_\mu$ parameters, cf. Eqs.~\eqref{eq:Krho}-\eqref{eq:Ksigma}, of the Fermi-Hubbard model~\eqref{eq:Microscopic_Lattice_Model} with DMRG. We consider a system of $ L = 41$ sites with open boundary conditions, and set the MPS bond dimension $D=200$, cutoff $10^{-12}$ and perform a DMRG calculation with 100 sweeps. $N/L$ denotes the filling fraction of the chain.}
    \label{fig:4}
\end{figure}

As a first step, we extract the LL parameters characterizing the fermionic environment. To this end, we consider an open chain geometry that is decoupled from the impurity and apply DMRG to the system at different filling fractions. The finite-size level repulsion, which we keep minimal but still well resolved within our numerical accuracy, ensures reliable access to the low-energy properties of the system. 
The LL parameters \mbox{$K_{\rho}$} and \mbox{$K_{\sigma}$} are obtained from the ground-state compressibility and spin susceptibility using Eqs.~\eqref{eq:Krho}-\eqref{eq:Ksigma}, evaluated for various filling fractions and Fermi-Hubbard interaction strengths. The extracted parameters for a lattice of \mbox{$41$} sites are shown in Fig.~\ref{fig:4}. By tuning both the density and the Fermi-Hubbard interaction, we access a broad interaction regime spanning \mbox{$0.2 \le K_{\nu} < 2.0$}. 
As expected, the LL parameters exhibit the characteristic dependence on the interaction strength: for a non-interacting gas they are unity, while repulsive interactions lead to \mbox{$K_{\rho}<1$} and \mbox{$K_{\sigma}>1$}, and attractive interactions yield \mbox{$K_{\rho}>1$} and \mbox{$K_{\sigma}<1$}. For most parameter regimes, we see that the LL parameters remain close to their non-intreracting value of unity. However for attractive Fermi-Hubbard interactions and low-fillings, a strongly interacting regime emerges where \mbox{$K_{\rho}\sim 2$}, \mbox{$K_{\rho}\sim 0.3$}.
These numerically extracted parameters allow for a direct and quantitative comparison with the LL theory and the RG flow analysis presented in Sec.~\ref{sec:RG}.

Having verified the LL parameters of the fermionic environment, we now couple the impurity to the system and vary the impurity interaction strengths at various points in the RG phase diagram. To identify the formation of a Kondo impurity in the DMRG simulations, we monitor two complementary observables. The first is the von Neumann entanglement entropy, \mbox{$S$}, defined in Eq.~\eqref{eq:von_neumann_entropy}, which quantifies the entanglement between the impurity and the fermionic environment. The second is the screening correlator, \mbox{$\mathcal{C}$}, defined in Eq.~\eqref{eq:screening_correlator}~\cite{Stocker2025}, which measures whether the spins of the LL atoms are aligned or antialigned with the impurity spin~\cite{Stocker2025}. Within the conventional Kondo paradigm, one expects the impurity to be nearly maximally entangled with the environment, \mbox{$S \approx \ln(2)$}, accompanied by near-perfect antiferromagnetic screening of the impurity spin, \mbox{$\mathcal{C} \approx -1$}.

\begin{figure*}
\centering
\includegraphics[width=17.6cm]{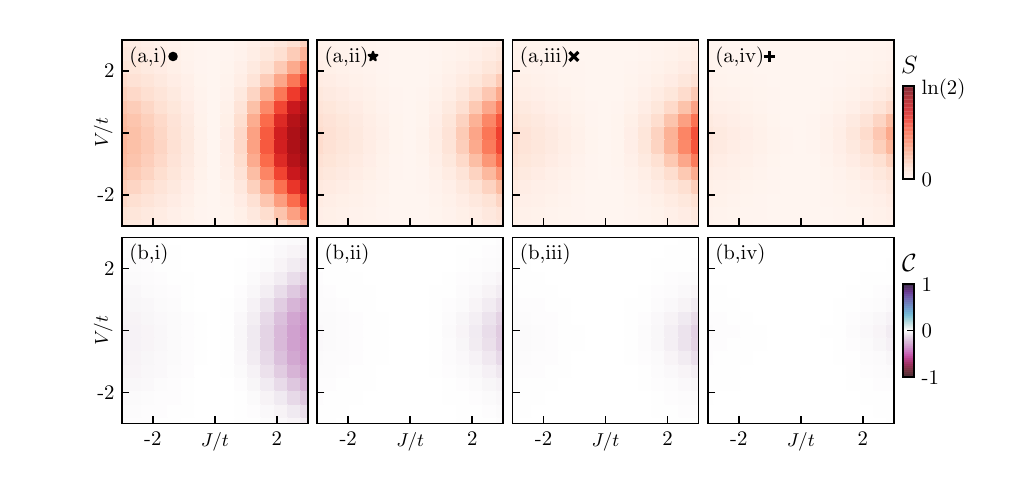}
\caption{\textit{DMRG results of impurity–LL hybridization and screening mechanisms.} We present results for (a) von Neumann entropy of the Yb impurity, cf. Eq.~\eqref{eq:von_neumann_entropy}, and (b) the chain screening correlator, cf. Eq.~\eqref{eq:screening_correlator}. The chosen parameters correspond to the white markers shown in Fig.~\ref{fig:4}, (i) for the $\mathbf{\cdot}$ marker $(K_\rho,K_\sigma) \approx (0.71,1.04)$,  (ii) for the ${\star}$ marker $(K_\rho,K_\sigma) \approx (1.03, 0.88)$,  (iii) for the ${\scriptstyle \times}$ marker $(K_\rho,K_\sigma) \approx (0.94, 0.67)$, and (iv) for the ${\scriptstyle +}$ marker $(K_\rho,K_\sigma) \approx (1.33, 0.38)$. These LL parameters correspond to the following filling fractions \mbox{$N/L$}, and Fermi-Hubbard Interactions \mbox{$U/t$}: (i) \mbox{$(N/L,U/t) = (20/41, 1.75)$}, (ii) \mbox{$(N/L,U/t) = (15/41, -0.45)$}, (iii) \mbox{$(N/L,U/t) = (14/41, -0.9)$}, (iv) \mbox{$(N/L,U/t) = (13/41, -1.75)$}.  
We also set chain length of $L=41$ atoms with open boundary conditions, MPS bond dimension  $D=500$, cutoff $10^{-12}$, and perform a DMRG calculation with $100$ sweeps. In (b) regions colored purple denote where the screening cloud is anti-aligned to the impurity, while regions colored teal are where the screening cloud is aligned with the impurity. We take the parameter regimes with both (a) large entanglement entropy and (b) an anti-aligned screening cloud as signatures of the Kondo effect.}
\label{fig:5}
\end{figure*}

The resulting zero-temperature observables are shown in Fig.~\ref{fig:5}, with the corresponding LL parameters explicitly indicated in Fig.~\ref{fig:2}(c). Depending on the location in parameter space, we observe qualitatively distinct behavior. In Fig.~\ref{fig:5}(i), we find a pronounced regime of strong impurity-environment entanglement, \mbox{$S \approx \ln(2)$}, which coincides with nearly perfect screening of the impurity spin by the LL, \mbox{$\mathcal{C} \to -1$}, for antiferromagnetic coupling \mbox{$J>0$}. This behavior constitutes a clear many-body signature of the Kondo effect. It has been reported Kondo should also appear for ferromagnetic interactions  \mbox{$J<0$}, in LLs~\cite{Lee1992}. We do not conclusively observe such an effect here, namely, we do observe finite entanglement, but the screening correlator is vanishingly small.

In Figs.~\ref{fig:5}(ii-iii), the parameter region supporting Kondo screening is significantly reduced. In the absence of potential scattering, the Kondo effect is confined to interaction strengths comparable to the kinetic energy scale, \mbox{$J/t \sim 2$}. The introduction of even moderate potential scattering \mbox{$V$} rapidly suppresses Kondo correlations. Comparing to the RG phase diagram in Fig.~\ref{fig:2}, Fig.~\ref{fig:5}(ii) is near the boundary where the magnetic backscattering \mbox{$\tilde{J}_{z,B}$} and potential scattering become marginal, while Fig.~\ref{fig:5}(iii) is in a regime in which the Kondo fixed point is unstable against potential scattering. The observed suppression of \mbox{$S$} and \mbox{$\mathcal{C}$} in the DMRG data is fully consistent with this RG prediction. We note that the reduction of the Kondo regime in the absence of potential scattering is partially influenced by finite-size effects.
Finally, Fig.~\ref{fig:5}(iv) exhibits a complete absence of Kondo signatures. While the generalized poor man’s scaling analysis in Fig.~\ref{fig:2}(a) would predict Kondo screening in this parameter regime, the DMRG results justifies the inclusion of the two particle-hole scattering terms, which  render the Kondo effect unstable. As a result, Kondo screening is suppressed even when potential scattering is absent. This demonstrates the crucial role of interaction-generated scattering processes beyond the simplest scaling analysis. 

\begin{figure*}
    \centering
    \includegraphics{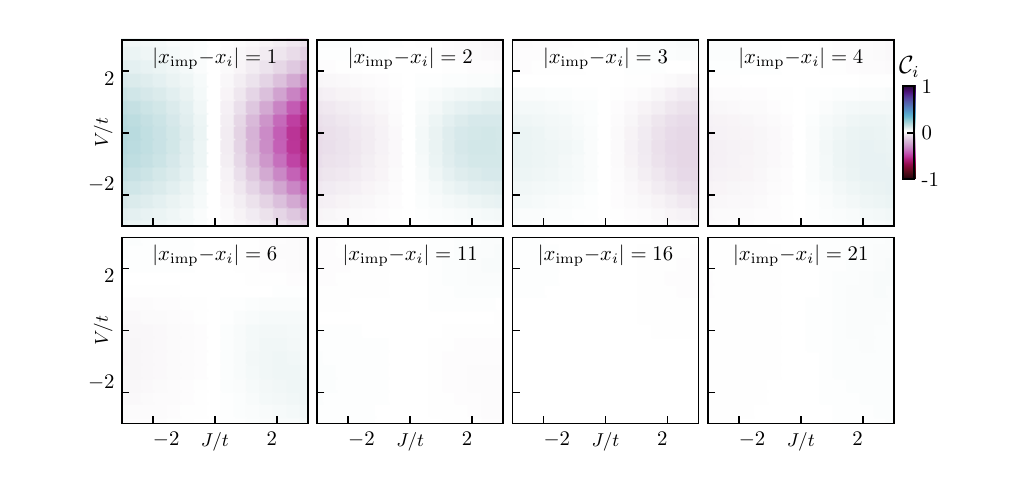}
    \caption{\textit{Spin screening correlator versus distance from the impurity.} We set $U = 2t$ and $N/L=20/41$, which corresponds to the $\cdot$ marker in Fig.~\ref{fig:3}(c) and the regime studied in Fig.~\ref{fig:5}(i). The other system's coupling and numerical DMRG parameters are as in Figs.~\ref{fig:5}(i). Purple (teal) regions are where the local screening correlator is anti-aligned (aligned) with the impurity of the spin. The local screening correlator is given by Eq.~\eqref{eq: total projected spin environment site} and is shown for various distances from the impurity.}
    \label{fig:6}
\end{figure*}

The observables in Fig.~\ref{fig:5} characterize global screening of the impurity by the fermionic environment. To distinguish between an extended Kondo cloud and a localized two-particle singlet, we further resolve the screening correlator spatially. 
In Fig.~\ref{fig:6}, we show the site-resolved correlator \mbox{$\mathcal{C}_i$}, introduced in Sec.~\ref{sec:DMRG}, for a representative subset of distances from the impurity site in the Kondo regime corresponding to Fig.~\ref{fig:5}(i). The correlator \mbox{$\mathcal{C}_i$} quantifies the contribution of each site to the screening process. We find that multiple sites in the LL chain contribute appreciably, \mbox{$\mathcal{C}_i \neq 0$}, for both antiferromagnetic and ferromagnetic couplings. For the parameters consistent with Fig.~\ref{fig:5}(i), the screening correlator has a correlation length of approximately \mbox{$5-6$} sites. This confirms that the observed screening arises from a genuinely many-body Kondo cloud rather than from a localized two-particle hybridization process. In other regimes where there is no Kondo effect, such as those in Fig.~\ref{fig:5}(iii-iv), we confirm that the screening correlator only has a finite value at the impurity site.
Taken together, these results establish that while potential scattering generically suppresses the Kondo effect, there exist well-defined interaction regimes in which Kondo screening persists, albeit with reduced robustness. By fixing the LL parameters and tuning the impurity couplings, we find quantitative agreement between the RG phase diagram and the exact DMRG ground state. This constitutes the second main result of this work.

\section{Renormalization group study of the Kondo effect with potential scattering}
\label{sec:RG}

In this section, we present the RG equations that we solve in Sec.~\ref{sec:pertRG} and show how we extract the Kondo temperature from the result. 

\subsection{Renormalization group equations}

Figs.~\ref{fig:2} and \ref{fig:3} were obtained using the zero-temperature Wilsonian RG formalism, in the weak coupling regime, calculated to second order in the cumulant expansion. We find that there are seven interactions needed to describe the RG flow in the weak coupling regime. These interactions are summarized in Tab~\ref{tab:1}. Five of these interactions correspond to the terms in the bare Hamiltonian, see Eq.~\eqref{eq:H}. There are two additional terms that are generated by the RG flow which are given by Eqs.~\eqref{eq:deltaH20}-\eqref{eq:deltaH02}.

The explicit calculation of the RG equations are presented in Appendix~\ref{app:RG}. Here, we solely present the resulting beta functions for the dimensionless interaction strengths of the system and compare their various limits to known results in the literature. The beta functions for the system read
\begin{widetext}
\begin{subequations}
\begin{align}
    \frac{d\eta}{d\ell} &= 2(1-\eta) 
    \frac{1
    }{K_{\sigma}^2}\left[\left(K_{\sigma}+\frac{(1-\eta)^2}{K_{\sigma}}\right)\tilde{J}_{\perp,F}^2 + \left(K_{\rho}+\frac{(1-\eta)^2}{K_{\sigma}}\right)\tilde{J}_{\perp,B}^2 \right] \ , \label{eq:RG_eta} \\
    \frac{d\tilde{J}_{z,B}}{d\ell} &= \left(1-\frac{K_{\rho}+K_{\sigma}}{2}\right) \tilde{J}_{z,B} +2K_{\rho}\tilde{V}_{2,0}\tilde{J}_{z,B}+2K_{\sigma}\tilde{V}_{0,2}\tilde{J}_{z,B} +4\frac{(1-\eta)^2}{K_{\sigma}} \tilde{J}_{\perp,F} \tilde{J}_{\perp,B} \ , \label{eq:RG_Sz_B}\\
    \frac{d\tilde{J}_{\perp,F}}{d\ell}&= \left(1-\frac{1}{2}\left(K_{\sigma}+\frac{(1-\eta)^2}{K_{\sigma}}\right)\right)\tilde{J}_{\perp,F} +2K_{\rho} \tilde{J}_{z,B}\tilde{J}_{\perp, B}+2 K_{\sigma} \tilde{V}_{0,2}\tilde{J}_{\perp, F} \ ,\label{eq:RG_Sperp_F}\\
    \frac{d\tilde{J}_{\perp,B}}{d\ell}&= \left(1-\frac{1}{2}\left(K_{\rho}+\frac{(1-\eta)^2}{K_{\sigma}}\right)\right)\tilde{J}_{\perp,B} +2K_{\sigma} \tilde{J}_{z,B}\tilde{J}_{\perp, F} +2 K_{\rho} \tilde{V}_{2,0}\tilde{J}_{\perp,B} \ , \label{eq:RG_Sperp_B}\\
    \frac{d\tilde{V}}{d\ell}&= \left(1-\frac{K_{\rho}+K_{\sigma}}{2}\right)\tilde{V} -2 K_{\rho}\tilde{V}_{2,0}\tilde{V} -2 K_{\sigma}  \tilde{V}_{0,2}\tilde{V} \ , \label{eq:RG_V_B} \\
    \frac{d\tilde{V}_{2,0}}{d\ell}&= (1-2K_{\rho})\tilde{V}_{2,0} +(K_{\rho}-K_{\sigma}) \tilde{V}^2 -(K_{\rho}-K_{\sigma})\tilde{J}_{z,B}^2 -2 \left(K_{\rho} - \frac{(1-\eta)^2}{K_{\sigma}}\right) \tilde{J}_{\perp,B}^2\label{eq:RG_V_20} \ ,  \\
    \frac{d\tilde{V}_{0,2}}{d\ell}&= (1-2K_{\sigma})\tilde{V}_{0,2} +(K_{\sigma}-K_{\rho})
    \tilde{V}^2-(K_{\sigma}-K_{\rho}) \tilde{J}_{z,B}^2 -2\left(K_{\sigma} - \frac{(1-\eta)^2}{K_{\sigma}}\right) \tilde{J}_{\perp,F}^2\label{eq:RG_V_02} \ ,
\end{align}
\label{eq:RG_equations_final}
\end{subequations}
\end{widetext}
where the definitions of the dimensionless interaction strengths are presented in Appendix~\ref{app:LL} and \mbox{$\ell = \ln(\Lambda/\mu)$} is a dimensionless measure of the ratio of the reduced ultraviolet cutoff \mbox{$\mu\sim T$} to the original ultraviolet cutoff \mbox{$\Lambda \sim v_F/a$}. In deriving Eq.~\eqref{eq:RG_equations_final}, we also neglected the forward potential scattering as it can be removed via a gauge transformation \cite{Giamarchi2003}. For simplicity, we also muted the dependence of the dimensionless coupling constants on \mbox{$\ell$} unless explicitly stated.

The set of beta functions presented in Eqs.~\eqref{eq:RG_equations_final} is a central result of this work. The beta functions for the magnetic scattering  Eqs.~\eqref{eq:RG_eta}-\eqref{eq:RG_Sperp_B} and potential scattering \eqref{eq:RG_V_B}-\eqref{eq:RG_V_02} presented here are qualitatively similar to those reported in Refs.~\cite{Lee1992,Schiller1997} which mapped the LL impurity scattering onto a two-dimensional Coulomb gas of electrical and magnetic charges. Our results differ from these previous works primarily by the inclusion of the LL parameters at second order, as well as a more systematic derivation of the beta functions for the leading irrelevant interactions, Eqs.~\eqref{eq:RG_V_20} and \eqref{eq:RG_V_02}. 

If we consider only the magnetic interactions and treat the deviations of the Luttinger parameters from unity as additional perturbations, our results reproduce the RG equations obtained from poor man's scaling~\cite{Furusaki1994}. Furthermore, Eqs.~\eqref{eq:RG_equations_final} agree with a Wilsonian approach to the Kondo problem, reported in Ref.~\cite{Kim2001}. However, they focused solely on the magnetic terms, and disregarded the effects of the potential scatterings. Finally, if one considers the RG flows of only the potential scattering terms, one recovers the Kane-Fisher model of impurity scatterings in a spin-1/2 LL~\cite{Kane1992,Kane1992b,Furusaki1993}.

An inspection of the full beta functions in Eq.~\eqref{eq:RG_equations_final} shows that these two paradigms, Kane-Fisher potential scattering~\cite{Kane1992,Kane1992b,Furusaki1993} and Kondo~\cite{Lee1992,Furusaki1994}, are coupled by the two particle-hole scattering terms \mbox{$\tilde{V}_{0,2}$} and \mbox{$\tilde{V}_{0,2}$}. Although these two particle-hole scattering terms are less relevant than the potential and magnetic scattering, they can ultimately alter the RG flows of the magnetic and potential scatterings, cf.~Figs.~\ref{fig:2} and~\ref{fig:7}. As such, our calculation presents the most complete description of the RG flow of the Kondo problem in a LL to second order in impurity scattering, and consistently incorporates the reduction of the Kondo temperature due to potential scattering.

\subsection{Extracting the Kondo temperature}
\label{sec: absence}

The Kondo temperature, \mbox{$T_K$}, is estimated as the temperature scale at which \mbox{$\tilde{J}(\ell^*\equiv \ln(\Lambda/T_K))=1$}, for any magnetic interaction \mbox{$\tilde{J}$}. First, consider the ideal case obtained from only the magnetic scattering terms. In this limit, the behavior of the Kondo temperature, \mbox{$T_K^0$}, depends on which term in the RG flow is dominant. For strongly repulsive interactions, \mbox{$K_{\rho} <1 $}, the fastest growing interaction is the backward scattering term. The growth rate is dominated by the scaling dimension and leads to a polynomial dependence on the magnetic scattering strength~\cite{Lee1992, Furusaki1994}
\begin{equation}
    T_K^0 \propto \Lambda \left(\frac{|J_{\perp,B}(0)|}{\Lambda}\right)^{\frac{2-K_{\rho}-K_{\sigma}^{-1}}{2}} \ .
    \label{eq:TKLL}
\end{equation}
Note that this result is valid for both ferromagnetic and antiferromagnetic interactions, unlike the case of Fermi liquids, where the RG flow suggests that the Kondo effect only exists for antiferromagnetic couplings~\cite{Anderson1970,Wilson1975}. 

In the weakly and non-interacting limit \mbox{$K_{\rho},K_{\sigma}\approx1$}, the preceding analysis breaks down. In that case, the RG flow reduces to the results obtained from Poor man's scaling~\cite{Anderson1970, Furusaki1994}. The dominant contribution to the RG flow in this case comes from the second-order terms and yields a Kondo temperature that depends on \mbox{$1/J_{\perp,B}$} exponentially
\begin{equation}
    \left.T_K^0\right|_{\text{FL}}\propto \Lambda e^{-2\pi v_F/J_{\perp,B}(0)}\, . 
    \label{eq:TKFL}
\end{equation}
When potential scattering terms are included, see Figs.~\ref{fig:2}(b) and (c), the Kondo effect can even be replaced with a strong coupling regime dominated by the potential scatterings in certain parameter regimes. In this case, it is the potential scattering terms that become larger than unity first. This is even true when \mbox{$\tilde{V}(0) = 0$}, as the magnetic interactions generate two particle-hole scattering that is relevant for \mbox{$K_{\rho}<1/2$} and \mbox{$K_{\sigma}<1/2$}.

For strictly non-interacting environments,  it is possible to obtain an estimate for the Kondo temperature in the presence of potential scattering. As discussed in Appendix~\ref{app:non_interacting}, the single particle wavefunctions in the presence of potential can be calculated analytically using scattering theory. Then employing a poor man's scaling approach leads one to a reduced Kondo temperature
\begin{equation}
    \left.\frac{T_K}{T_K^0}\right|_{\text{non int.}} = e^{-\pi V^2/(Jv_F)}\ .
\end{equation}

\section{Hybridization and screening of the Yb impurity}
\label{sec:DMRG}

In our numerical analysis, we study the ground state of the Fermi–Hubbard model in Eq.~\eqref{eq:Microscopic_Lattice_Model} using DMRG within the matrix product state framework~\cite{white1992,schollwock2011}, which provides a variational approach to ground-state optimization. For our simulations, we choose a system's dimension of $L=41$ and MPS truncation parameter (bond dimension) $D=500$, as these yield convergent results. In this section, we introduce the correlation functions and order parameters used to analyze the resulting quantum many-body ground state obtained from our DMRG calculations.

\subsection{Impurity--LL hybridization}

To quantify the hybridization between the impurity and the LL, we consider the reduced density matrix of the impurity 
\begin{equation}
    \rho_{imp} = \Tr_{LL}\left(\rho\right),
\end{equation}
where the total system composed of the impurity and the LL is described by 
\begin{equation}
    \rho = \ket{\psi}\bra{\psi}.
\end{equation}
In experiments, the impurity’s density matrix, $\rho_{imp}$, can be accessed through a tomography procedure, which consists of a diabatic decoupling of the environment~\cite{Stocker2022}. Thus, one can read out different observables to estimate the amount of system-environment coupling and the many-body coupling mechanism. For illustration, we consider the von Neumann entropy of the impurity atom
\begin{equation}
    S = -\Tr\left[\rho_{imp}\ln\left(\rho_{imp}\right)\right], \label{eq:von_neumann_entropy}
\end{equation}
which serves as a measure of the entanglement between the impurity and the LL. For an impurity fully that is decoupled from the environment, we have \mbox{$S=0$}. Hybridization between the impurity atom and the fermionic environment leads to \mbox{$S>0$}, with a maximum value corresponding to a completely mixed impurity state, \mbox{$S=\ln(2)$}.

\subsection{Impurity--LL screening}

While the von Neumann entropy quantifies the amount of entanglement between the impurity and environment, it does not reveal how the LL screens the impurity. To address this, we employ a correlator recently proposed in Ref.~\cite{Stocker2025}, which captures the spin polarization of the environment conditional on the spin state of the impurity. For its evaluation, we define the projection operators
\begin{equation}\label{eq: projector operators spin subhilbert spaces}
	{P}^{\sigma}_{imp}\equiv \ket{\sigma}\bra{\sigma}_{imp}\otimes{I}_\text{LL},
\end{equation}
where ${I}_\text{LL}$ is the identity operator acting on the LL. Using these operators, we compute the correlator
\begin{equation}\label{eq:screening_correlator}
\mathcal{C} = \sum_\sigma\frac{1}{\bra{\sigma}{S}_{imp}^z\ket{\sigma}}\bra{\psi} {P}^{\sigma}_{imp}{S}_{LL}^z{P}^{\sigma}_{imp}\ket{\psi},
\end{equation}
where ${S}_{imp}^z$ (${S}_{LL}^z$) denotes the $z$-component of the spin operator for the impurity (LL). The correlator takes the value $\mathcal{C}=1$ ($-1$) if the total spin of the environment is aligned (antialigned) with the impurity spin. If there is no alignment, the correlator returns $\mathcal{C} = 0$. Along the MPS chain, the correlator of the environment is the sum over the correlator of each site
\begin{equation}
	\mathcal{C} = \sum_{i} \mathcal{C}_{i} \ , \label{eq: spin projected space environment as sum of wilson chain sites}
\end{equation}
where
\begin{equation}\label{eq: total projected spin environment site}
\mathcal{C}_{i} =  \sum_{\sigma}\frac{1}{\bra{\sigma}{S}^z\ket{\sigma}}\bra{\psi}P^{\sigma} S_{i}^{z}{P}^{\sigma}\ket{\psi} ,
\end{equation}
is the site-resolved correlator of the $i^\mathrm{th}$ site of the LL chain. 
Since we place the impurity at the center of our LL, the screening correlator is symmetric about the impurity site, and only depends on the distance away from the impurity. This is shown in Fig.~\ref{fig:6}.
We take the combined presence of strong entanglement entropy and a strong anti-aligned screening cloud as the confirmation of Kondo correlations in the systems

\section{Conclusion and outlook}
\label{sec:conclusions}

In this work, we show that the effect of non-magnetic impurity scattering depends sensitively on the correlations in the one-dimensional Fermionic environment. We identified clear parameter regimes, see Fig.~\ref{fig:2}, where the Kondo effect persists at asymptotically low temperatures, in spite of the non-magnetic impurity scattering being a relevant perturbation. This is in contrast to other parameter regimes where the non-magnetic impurity scattering can dominate, and form non-magnetic orders. These results highlight how correlations in the fermionic environment can potentially protect the Kondo phase, facilitating its observation in cold atoms.

Within a continuum field-theoretic description, our renormalization-group analysis reveals that impurity-generated two particle-hole scattering processes are essential for coupling magnetic and potential impurity scattering. These terms are often neglected in the literature~\cite{Lee1992,Furusaki1994,Kim2001,Giamarchi2003}, an approximation that is only valid for weak potential scattering. Once the potential scattering strength becomes comparable to the magnetic exchange, the generated interactions lead to a strong and experimentally relevant suppression of the Kondo temperature. This identifies a previously overlooked mechanism by which correlations hinder Kondo screening in one dimension.
Conceptually, our work closes an important gap in the theoretical description of Kondo impurities in correlated environments. By treating magnetic exchange, potential scattering, and impurity-generated two particle-hole scattering on equal footing within a controlled RG framework, we identify the minimal ingredients required to determine whether Kondo screening survives or is preempted. This yields clear theoretical constraints for the onset of the Kondo effect in interacting atomic gases and delineates the regimes where simplified treatments break down.

These analytical results are quantitatively supported by density matrix renormalization group calculations of the microscopic lattice model. By directly accessing the many-body ground state, we observe a clear crossover from a Kondo regime, characterized by near-maximal impurity entanglement and collective many-body screening, to a weakly interacting regime in which the impurity and the Luttinger liquid are only weakly entangled. The agreement between the RG phase diagram and the exact numerical results provides a coherent and predictive picture of how potential scattering destabilizes Kondo physics in one dimension.

From an experimental perspective, our findings are directly relevant to ongoing efforts to realize Kondo physics in ultracold atomic gases. The effects of potential scattering can be compensated by locally tuning the impurity chemical potential, for example using optical tweezers. This leads to an effective scattering potential of strength: \mbox{$V_{eff} = V-\mu_0$} where \mbox{$\mu_0$} is the local chemical potential for the impurity atom. Fine tuning the local chemical potential enables access to an effectively ideal Kondo regime and provides a controlled knob for systematically exploring the interplay between magnetic and non-magnetic impurity scattering. 
More broadly, cold-atom realizations of the Kondo effect enable probes of impurity physics that go beyond transport-based diagnostics in condensed matter systems. State-selective imaging offers direct access to the impurity state and its entanglement with the environment, while local two body losses on the spin impurity can further promote Kondo \cite{Stefanini2025}. This opens the door to studying the formation dynamics of the Kondo cloud, realizing Kondo lattice models, and simulating open quantum systems coupled to correlated many-body environments. Atomic platforms thus provide a powerful setting for exploring impurity physics in regimes that remain difficult to access in solid-state systems.

\begin{acknowledgments}
We thank F. Scazza, M. Aidelsburger, R. Kroeze, and T. Giamarchi for fruitful discussions. We acknowledge funding from the Deutsche Forschungsgemeinschaft (DFG) through project number 449653034 and through the research unit FOR5688, project number 521530974. We also acknowledge support from the Swiss National Science Foundation (SNSF) through NCCR SPIN and through project 190078. The numerical MPS implementations are based on the \textsc{ITensors.jl} library~\cite{Fishman2022}.
\end{acknowledgments}

\section*{APPENDICES}

\appendix

\section{Review of Luttinger liquid theory}
\label{app:LL}

In this Appendix, we review the key aspects of Luttinger liquid (LL) theory relevant to the renormalization group (RG) analysis using the notation of Ref.~\cite{vonDelft1998}. The starting point is the total Hamiltonian in the continuum limit, Eq.~\eqref{eq:H}. The individual terms in Eq.~\eqref{eq:H} are given by Eqs.~\eqref{eq:HLL}-\eqref{eq:Uex_pm}. The excitations in the absence of the impurity are bosonic fluctuations in either the density (\mbox{$\nu = \rho$}) or spin (\mbox{$\nu = \sigma$}). These fluctuation are controlled by the bosonic fields \mbox{$\theta_{\nu}$} and \mbox{$\varphi_{\nu}$} which have the commutation relations stated in Eq.~\eqref{eq:commutator}. The fermionic operators are related to the bosonic ones via Eq.~\eqref{eq:bosonized_fermion}.

In order to evaluate the normal ordered products of fermionic operators in Eqs.~\eqref{eq:HLL}-\eqref{eq:Uex_pm} it is necessary to take care of (i) the commutation relations of the bosonic fields, and (ii) the anticommutation relations of the Klein factors \mbox{$\left\lbrace F_{r,s}, F^{\dagger}_{r',s'}\right\rbrace = 2\delta_{r,r'}\delta_{s,s'}$}.

The commutation relations of the bosonic fields can be correctly accounted for by using a point-splitting method to evaluate the normal ordering of the operators~\cite{Haldane1981,vonDelft1998}. In order to account for the exchange properties of the Klein factors, we note that any bilinear combination of Klein factors can be represented by the Pauli sigma matrices \mbox{$\tau_{x,y,z}$}~\cite{Egger1996,Egger1998}
\begin{align}
    F_{r,s}^{\dagger}F_{\bar{r},s} &= s \tau_z & F_{r,s}^{\dagger}F_{r,\bar{s}} &= i s \tau_y & F_{r,s}^{\dagger}F_{\bar{r},\bar{s}} &= \tau_x.
    \label{app:Klein_factors}
\end{align}
In terms of the bosonic variables and the \mbox{$\tau_{x,y,z}$} matrices,  the impurity scattering terms then read
\begin{subequations}
\begin{align}
    \delta H_{V} &= -\frac{4V}{2\pi a} \tau_z \sin(\sqrt{2}\varphi_{\rho}(0))\sin(\sqrt{2}\varphi_{\sigma}(0)) \ , \label{app:deltaH_V} \\
    \delta H_{z,F} &= \frac{J_{z,F}}{2\pi a} S_z \sqrt{2}a \partial_x \varphi_{\sigma}(0) \label{app:deltaH_ZF} \ , \\
    \delta H_{z,B} &= 2\frac{J_{z,B}}{2\pi a} S_z\tau_z \cos(\sqrt{2}\varphi_{\rho}(0))\cos(\sqrt{2}\varphi_{\sigma}(0)) \label{app:deltaH_ZB} \ , \\
    \delta H_{s,F} &= \frac{J_{\perp,F}}{2\pi a}S_{\bar{s}} i s \tau_ye^{i\sqrt{2}s \theta_{\sigma}(0)} \cos(\sqrt{2}\varphi_{\sigma}(0)) \label{app:deltaH_sF} \ , \\
    \delta H_{s,B} &= \frac{J_{\perp,B}}{2\pi a}S_{\bar{s}} \tau_x e^{i\sqrt{2}s \theta_{\sigma}(0)} \cos(\sqrt{2}\varphi_{\rho}(0)) \ . \label{app:deltaH_sB}
\end{align}
\end{subequations}
We will also include in the Hamiltonian the leading terms generated by the RG flow which describe processes that involve the creation of two particle-hole pairs
\begin{subequations}
\begin{align}
        \delta H_{2,0} &= \frac{1}{2} \frac{V_{2,0}}{(2\pi (a)^2} \cos(2\sqrt{2}\varphi_{\rho}(0)) \ , \label{app:deltaH_20} \\
    \delta H_{0,2} &= \frac{1}{2} \frac{V_{0,2}}{(2\pi (a)^2} \cos(2\sqrt{2}\varphi_{\sigma}(0)) \ . \label{app:deltaH_02}
\end{align}
\end{subequations}
As described in the main text, such terms play a crucial role in determining how the potential scattering influences the RG flow of the magnetic couplings. Although all the magnetic terms are initially equal, they evolve differently under the RG flow. Hence, we treat the coupling constants for each of the above terms as different variables.

An important simplification is to remove the forward scattering terms by means of a gauge transformation~\cite{Giamarchi2003}. Consider applying the following gauge transformation to the Hamiltonian
\begin{equation}
    U_{\sigma} = e^{-i s \frac{K_{\sigma}}{v_{\sigma}} \frac{J_{z,F}}{2\pi} \sqrt{2} \theta_{\sigma}(0)} \ .
    \label{eq:gauge_transformation}
\end{equation}
This results in
\begin{align}
    H_{LL} &\to H_{LL} - \delta H_{z,F}\ , \nonumber \\
    \delta H_{s,F} &\to \frac{J_{\perp,F}}{2\pi a}S_{\bar{s}}i s \tau_y e^{i \sqrt{2}s(1-\eta)\theta_{\sigma}(0)} \cos(\sqrt{2}\varphi_{\sigma}(0)) \ , \nonumber \\
    \delta H_{s,B} &\to \frac{J_{\perp,B}}{2\pi a}S_{\bar{s}}i s \tau_y e^{i \sqrt{2}s(1-\eta)\theta_{\sigma}(0)} \cos(\sqrt{2}\varphi_{\rho}(0)) \ .
\end{align}
Here, we denote \mbox{$\eta =\frac{K_{\sigma}}{v_{\sigma}} \frac{J_{z,F}}{2\pi}$}. As a result, we can eliminate the effects of the forward magnetic scattering along \mbox{$z$} by introducing the parameter \mbox{$\eta$}. It is possible to perform a similar gauge transformation to remove the forward potential scattering. The result is trivial as no additional phase factors are introduced in the Hamiltonian. Thus, we are free to ignore the forward potential scattering in our analysis.

To facilitate our RG discussion, we define dimensionless parameters in units of the ultraviolet cutoff, \mbox{$\Lambda$}, in momentum space
\begin{align}
    \tilde{V} &= \frac{4V}{2\pi a \Lambda} & \tilde{J}_i &= \frac{J_i}{2\pi a \Lambda}  & \eta = \frac{K_{\sigma}}{v_{\sigma}} \tilde{J}_{z,F} a\Lambda \nonumber \ , \\
    \tilde{V}_{2,0} &= \frac{V_{2,0}}{(2\pi a)^2 \Lambda} & \tilde{V}_{0,2} &= \frac{V_{2,0}}{(2\pi a)^2 \Lambda} 
\end{align}
where \mbox{$J_i = \lbrace J_{\perp,F}, \ J_{\perp,B}, \ J_{z,B}\rbrace$}.

\section{Outline of Renormalization Group Approach}
\label{app:RG}

\subsection{Wilsonian RG with operators}
In this Appendix, we outline the Wilsonian renormalization group procedure up to second order in the cumulant expansion. We consider the finite temperature partition function
\begin{equation}
    Z = Tr[e^{-\beta H}] \ ,
\end{equation}
where \mbox{$\beta = 1/T$} is the inverse temperature and we have set Boltzmann's constant to $k_B = 1$. The total Hamiltonian is the sum over the quadratic LL Hamiltonian \mbox{$H_{LL}$}, Eq.~\eqref{eq:HLL}, and the total perturbation \mbox{$\delta H$}, Eq.~\eqref{eq:H}. The total perturbation is the sum of all the magnetic and potential scatterings in Eqs.~\eqref{eq:Udir}-\eqref{eq:Uex_pm}, as well as the two particle-hole processes in Eqs.~\eqref{eq:deltaH20}-\eqref{eq:deltaH02}.

The RG is implemented by separating the bosonic fluctuations into slow and fast modes with respect to a reduced cutoff scale \mbox{$\mu$} 
\begin{align}
    \varphi_{\nu}^s(x,\tau) &= \sum_{|\omega_n|<\mu} \varphi_{\nu}(x,\omega_n)e^{-i\omega_n \tau} \ , \nonumber \\
    \varphi_{\nu}^f(x,\tau) &= \sum_{\mu <|\omega_n|<\Lambda} \varphi_{\nu}(x,\omega_n)e^{-i\omega_n \tau} \ ,
    \label{app:slow_fast_sep}
\end{align}
and similarly for the fields \mbox{$\theta_{\nu}(x,\tau)$}.

In general, we can decompose the impurity scattering terms into a slow component, a fast component, and a coupling term \mbox{$\delta H = \delta H_s + \delta H_f + \delta H_c$}. We then trace out the fast components to obtain an effective Hamiltonian for the slow degrees of freedom. Formally, this is achieved by rewriting the partition function
\begin{align}
    Z &=  Tr_s\left[ e^{-\beta H_s} Tr_f\left[e^{-\beta H_{LL,f}} U(\beta)\right]\right] \ , \nonumber \\
    &= Z_fTr_s\left[ e^{-\beta H_s} \left\langle U(\beta) \right\rangle_f\right]  \ .
\end{align}
Here, we define \mbox{$Z_f$} as the partition function for the fast modes, and 
\begin{equation}
    U(\beta) = T_{\tau}\exp\left[-\int_0^{\beta}d\tau \delta H_c(\tau)\right] \ ,
\end{equation}
with \mbox{$\delta H_c(\tau)= e^{H_{LL,f}\tau}\delta H_c e^{-H_{LL,f}\tau}$}. We also define the average over the fast modes to be \mbox{$\langle \cdot \rangle_f = Tr_f\left[e^{-\beta H_f} \cdot \right]/Z_f$}.

In this work, we obtain the effective Hamiltonian to second order. This requires perturbatively evaluating \mbox{$\left\langle U(\beta) \right\rangle_f$} and then re-exponentiating the result using the cumulant expansion. This procedure yields an effective Hamiltonian for the slow modes given by
\begin{align}
    H_{s}^{eff} &= H_s + \frac{1}{\beta}\int_0^{\beta} d\tau \langle \delta H_c(\tau) \rangle_f -\frac{1}{2\beta } \int_0^{\beta} d\tau \int_0^{\beta} d\tau' \nonumber \\
    &\left[\langle \delta H_c(\tau)\delta H_c(\tau')\rangle_f - \langle \delta H_c(\tau)\rangle_f \langle \delta H_c(\tau')\rangle_f\right] \ .
    \label{app:effective_Hs}
\end{align}
At first order, the correction to the Hamiltonian is just the average of \mbox{$\delta H_c(\tau)$} over the fast modes, while at second order it is related to the connected part of the correlation function of \mbox{$\delta H_c(\tau)$}. 
In the remaining calculations, we treat the zero-temperature limit with \mbox{$\beta = \infty$}. Below, we provide explicit examples of how to obtain the RG flow equations to second order.

\subsection{First order calculation}

First, let us consider integrating out the fast-degrees of freedom to first order. This can be highlighted, for example, by examining \mbox{$\delta H_{V}$}. The explicit form of the interaction in terms of the bosonic fields is shown in Eq.~\eqref{app:deltaH_V}. Separating the system into slow and fast modes gives the following contribution to Eq.~\eqref{app:effective_Hs}
\begin{align}
    I_V &= -\tilde{V}\int_0^{\infty} d\tau  \tau_z  \nonumber \\
    &\left\langle\left[\sin(\sqrt{2}\varphi_{\rho}^s(\tau)+\sqrt{2}\varphi_{\rho}^f(\tau))\right]\times \left[\rho \to \sigma\right]\right\rangle_f \ , \nonumber \\
    &= -\tilde{V} \int_0^{\infty} d\tau \tau_z \left[\sin(\sqrt{2}\varphi_{\rho}^s(\tau)) \langle \cos(\sqrt{2}\varphi_{\rho}^f(\tau))\rangle_f \right . \nonumber \\
    &\left. + \cos(\sqrt{2}\varphi_{\rho}^s(\tau)) \langle \sin(\sqrt{2}\varphi_{\rho}^f(\tau))\rangle_f\right] \times \left[\rho \to \sigma\right]\,,
    \end{align}
where we suppress the position dependence of the fast modes, as all the couplings are evaluated at \mbox{$x=0$}.

As a next step, we note that the fast field propagator at \mbox{$x=0$} has the form~\cite{Fisher1985}
\begin{align}
    \langle \varphi_{\nu}^f(\tau)\varphi_{\nu}^f(\tau') \rangle &= K_{\nu} \tilde{G}(\tau-\tau') \ , \nonumber \\
    \langle \theta_{\nu}^f(\tau)\theta_{\nu}^f(\tau') \rangle &= K_{\nu}^{-1} \tilde{G}(\tau-\tau') \ , \nonumber \\
    \langle \theta_{\nu}^f(\tau)\varphi_{\nu}^f(\tau') \rangle &= 0 \ ,
    \label{app:propagators}
\end{align}
where \mbox{$\tilde{G}(\tau)$} is a short-ranged function
\begin{equation}
    \tilde{G}(\tau) =\begin{cases}
        \frac{1}{2}\ln\left(\frac{\Lambda}{\mu}\right) & \Lambda \tau \ll 1 \\
        \frac{1}{2}K_0(\mu\tau) & \Lambda \tau \gg 1 \ .
    \end{cases}
    \label{app:G}
\end{equation}
From Eqs.~\eqref{app:propagators}-\eqref{app:G}, we can evaluate the averages over the fast fields
\begin{align}
    I_V = -\tilde{V} \int_0^{\infty} d\tau & \tau_z \left(\frac{\Lambda}{\mu}\right)\left(\frac{\mu}{\Lambda}\right)^{\frac{K_{\rho}+K_{\sigma}}{2}} \nonumber \\
    &\sin(\sqrt{2}\varphi_{\rho}^s(\tau)) \sin(\sqrt{2}\varphi_{\sigma}^s(\tau) \ ,
    \label{eq:IV_first order}
\end{align}
 where we rescale the time coordinate \mbox{$\tau \to \tau \Lambda/\mu$} in order to return the cutoff to its original value. The quantity $I_V$ in Eq.~\eqref{eq:IV_first order}, resembles a correction to the coupling constant $\tilde{V}$. Upon re-exponentiating this term and setting \mbox{$\mu = \Lambda - \delta \Lambda$}, we obtain, to leading order O($\tilde{V}$)
\begin{equation}
    \tilde{V}(\Lambda-\delta \Lambda) = \tilde{V} + \left(1- \frac{K_{\rho}+K_{\sigma}}{2}\right)\tilde{V} \frac{\delta \Lambda}{\Lambda} \ .
\end{equation}
This allows us to finally write the change in the dimensionless coupling \mbox{$\tilde{V}$} as:
\begin{equation}
    \frac{d\tilde{V}(\Lambda)}{d\ell} = - \Lambda \frac{d\tilde{V}(\Lambda)}{d\Lambda} = \left(1- \frac{K_{\rho}+K_{\sigma}}{2}\right)\tilde{V} \ .
\end{equation}
This is the expected result obtained previously in Ref.~\cite{Kane1992}.
Similar calculations can be done to obtain the RG flow of the remaining perturbations to first order. 

\subsection{Second order calculation}

The calculation of the effective Hamiltonian at second order requires the calculation of all possible pairs of perturbations. Here, we showcase the procedure with two representative examples. Consider the second order term involving two forward magnetic interactions given by Eq.~\eqref{app:deltaH_sF}
\begin{align}
    &I_{FF} = \sum_{s}\int_0^{\infty} d\tau \int_0^{\infty} d\tau' \ \tilde{J}_{\perp,F}^2 is\tau_y i \bar{s} \tau_y S_s S_{\bar{s}} \nonumber \\
    &\left\langle e^{i s \sqrt{2}(1-\eta)\theta_{\sigma}(\tau)}e^{-i s \sqrt{2}(1-\eta)\theta_{\sigma}(\tau')} \right\rangle_f\nonumber \\
    &\left\langle\cos(\sqrt{2}\varphi_{\sigma}(\tau)) \cos(\sqrt{2}\varphi_{\sigma}(\tau'))\right\rangle_f \ .
\end{align}
For this term, the Klein factors simplify to unity. Explicitly separating the bosonic fields into slow and fast modes and integrating over the fast modes gives
\begin{align}
    I_{FF} &= \nonumber \\
    &\sum_s \int_0^{\infty} d\tau \int_0^{\infty} d\tau'  \tilde{J}_{\perp,F}^2  S_s S_{\bar{s}} \left(\frac{\Lambda}{\mu}\right)^2 \left(\frac{\mu}{\Lambda}\right)^{K_{\sigma} + \frac{(1-\eta)^2}{K_{\sigma}}}  \nonumber \\
    &e^{i s \sqrt{2}(1-\eta)(\theta_{\sigma}(\tau)-\theta_{\sigma}(\tau'))} e^{-\frac{2(1-\eta)^2}{K_{\sigma}}\tilde{G}(\tau-\tau')} \nonumber \\
    &\frac{1}{2}\left[\cos(\sqrt{2}(\varphi_{\sigma}(\tau)+ \varphi_{\sigma}(\tau')) e^{-2K_{\sigma}\tilde{G}(\tau-\tau')} \right. \nonumber \\
    &\left. + \cos(\sqrt{2}(\varphi_{\sigma}(\tau) - \varphi_{\sigma}(\tau')) e^{2K_{\sigma}\tilde{G}(\tau-\tau')}\right] \ .
    \label{app:intermediate_secondorder_1}
\end{align}
To evaluate Eq.~\eqref{app:intermediate_secondorder_1} we first subtract off the disconnected piece of the correlation function, see Eq.~\eqref{app:effective_Hs}, and observe that the function \mbox{$\tilde{G}(\tau-\tau')$} is extremely short ranged. 

We then move to a center of mass and relative time coordinates, \mbox{$T = (\tau+\tau')/2$} and \mbox{$\Delta \tau = \tau-\tau'$}, and perform a gradient expansion in \mbox{$\Delta \tau$}. The relevant terms are
\begin{align}
    I_{FF} &\approx \nonumber \\
    &\sum_s \int_0^{\infty} dT \int_{-\infty}^{\infty} d\Delta \tau  \tilde{J}_{\perp,F}^2  S_s S_{\bar{s}} \nonumber \\
    &\left(\frac{\Lambda}{\mu}\right)^2 \left(\frac{\mu}{\Lambda}\right)^{K_{\sigma} + \frac{(1-\eta)^2}{K_{\sigma}}}  \nonumber \\
    &\Big\{(1+i s \sqrt{2}(1-\eta)\Delta \tau \partial_T \theta_{\sigma}(T)]) e^{-\frac{2(1-\eta)^2}{K_{\sigma}}\tilde{G}(\Delta(\tau)} \nonumber \\
    &\frac{1}{2}\left[\cos(2\sqrt{2}(\varphi_{\sigma}(T) )e^{-2K_{\sigma}\tilde{G}(\Delta \tau)} +e^{2K_{\sigma}\tilde{G}(\Delta \tau)}\right]\nonumber \\
    & -\cos(2\sqrt{2}\varphi_{\sigma}(T))-1\Big\}.
    \label{app:intermediate_secondorder_2}
\end{align}
At this stage, it is possible to perform the summation over the index \mbox{$s$}. One can show that
\begin{align}
    \sum_s S_s S_{\bar{s}} &=S_+S_- + S_- S_+  = 2(S^2-S_z^2) = 1\ , \\
    \sum_s s S_s S_{\bar{s}} &= \left[S_+,S_-\right] = 2S_z\ .
\end{align}
Combining this result with the integration over the relative time-coordinate yields a contribution to \mbox{$\delta H_{0,2}$} and \mbox{$\delta H_{z,F}$}. The correction to the RG flow of \mbox{$\tilde{V}_{0,2}$} can be immediately read from Eq.~\eqref{app:intermediate_secondorder_2} by looking at the coefficients of the terms proportional to \mbox{$\cos(2\sqrt{2}\varphi_{\sigma}(T)$}. These terms lead to the following contribution to the RG flow of \mbox{$\tilde{V}_{0,2}$}.
\begin{equation}
    \frac{d\tilde{V}_{0,2}}{d\ell} \propto -2\left(K_{\sigma} + \frac{(1-\eta)^2}{K_{\sigma}}\right)\tilde{J}_{\perp,F}^2 \ .
\end{equation}

To see that we generate a perturbation of the form \mbox{$\delta H_{z,F}$} and a renormalization group flow of the parameter \mbox{$\eta$}, we note that the equation of motion for the field \mbox{$\theta_{\sigma}(T)$} is: 
\begin{equation}
    i \partial_T \theta_{\sigma}(T) = \frac{v_{\sigma}}{K_{\sigma}} \partial_x \varphi_{\sigma}(T) \ .
\end{equation}
This term has the same form as the forward magnetic scattering along \mbox{$z$} with strength
\begin{equation}
    \tilde{J}_{z,F}(\Lambda-\delta \Lambda) = \frac{2(1-\eta)}{K_{\sigma}^2}\left[K_{\sigma} + \frac{(1-\eta)^2}{K_{\sigma}}\right]\tilde{J}_{\perp,F}^2 \ .
\end{equation}
As was done in Appendix~\ref{app:LL}, we can remove this new forward scattering term by means of a gauge transformation. This leads to a term in the RG flow equations for the parameter \mbox{$\eta$}
\begin{equation}
    \frac{d\eta}{d\ell} \propto \frac{2(1-\eta)}{K_{\sigma}^2}\left[K_{\sigma} + \frac{(1-\eta)^2}{K_{\sigma}}\right]\tilde{J}_{\perp,F}^2 \ .
\end{equation}

To illustrate the importance of the Klein factors, we consider the second order term involving both forward and backward magnetic scatterings, Eqs.~\eqref{app:deltaH_sF}-\eqref{app:deltaH_sB}
\begin{align}
    I_{FB}&=\sum_s \int_0^{\infty}d\tau \int_0^{\infty} d\tau' \tilde{J}_{\perp,F}\tilde{J}_{\perp,B} i s \tau_y \tau_x S_{s} S_{\bar{s}} \nonumber \\
    &\left\langle e^{i s \sqrt{2}(1-\eta)\theta_{\sigma}(\tau)}e^{-i s \sqrt{2}(1-\eta)\theta_{\sigma}(\tau')} \right\rangle_f\nonumber \\
    &\langle\cos(\sqrt{2}\varphi_{\sigma}(\tau))\rangle_f\langle\cos(\sqrt{2}\varphi_{\rho}(\tau'))\rangle_f \ .
\end{align}
Integrating out the bosonic fast modes proceeds as in the previous example and will not be shown explicitly. Here, we emphasize that at second order both permutations of these perturbations appear in the evaluation of Eq.~\eqref{app:effective_Hs} and as such
\begin{align}
    I_{FB} &\propto \sum_s \left[i s\tau_y \tau_x S_s S_{\bar{s}} + \tau_x i s\tau_y S_{\bar{s}} S_{s}\right] \nonumber \\
    &\propto \sum_s i s \left[\tau_y,\tau_x\right] S_s S_{\bar{s}} \nonumber \\
    &\propto 4 \tau_zS_z \ .
\end{align}
Thus, \mbox{$I_{FB}$} generates a contribution to the magnetic backscattering along \mbox{$z$}. An explicit calculation yields the following term in the RG flow
\begin{equation}
    \frac{d\tilde{J}_{z,B}}{d\ell} \propto \frac{4(1-\eta)^2}{K_{\sigma}}\tilde{J}_{\perp,F}\tilde{J}_{\perp,B} \ .
\end{equation}

\subsection{Example of RG flow}

\begin{figure*}
    \centering
    \includegraphics{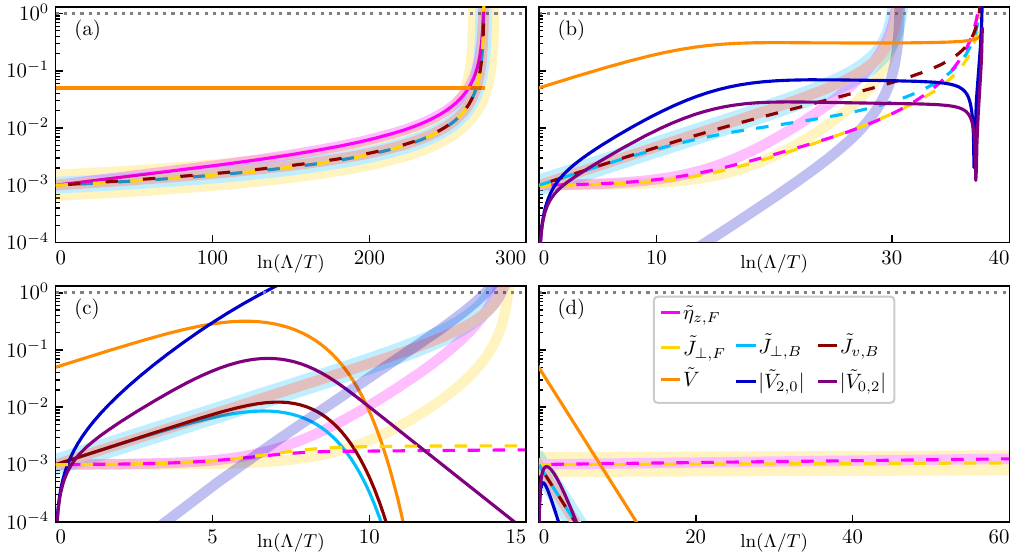}
    \caption{\textit{RG flows of the magnetic and potential impurity interactions.} Thick transparent lines are the RG flows in the absence of potential scattering, which corresponds to \mbox{$\tilde{V}(0) = 0$}. The initial values of the magnetic interactions are chosen to be: \mbox{$\tilde{J}(0) = 10^{-3}$}. Thin lines are for RG flows in the presence of potential scattering, which we initially set to \mbox{$\tilde{V}(0)=5\times10^{-2}$}. We set (a) \mbox{$K_{\rho}= 1.0$}, (b) \mbox{$K_{\rho}=0.7$}, (c) \mbox{$K_{\rho}=0.2$}, and (d) \mbox{$K_{\rho}=2.0$}. We consider a SU(2) invariant gas with \mbox{$K_{\sigma}=1$}, and initiate the RG flows \mbox{$\tilde{J}(0)=10^{-2}$}. We use dashed alternating thin lines and different thicknesses for opaque lines when results overlap and would otherwise be indistinguishable.}
    \label{fig:7}
\end{figure*}

In Fig.~\ref{fig:7}, we show some characteristic trajectories of the RG flows in the presence and absence of potential scatterings for the SU(2) invariant case, \mbox{$K_{\sigma}=1$}. Here, we show the RG flows for (a) the non-interacting point, \mbox{$K_{\rho}=1.0$}, (b) a point where the potential scattering is relevant, while the two particle-hole scattering is irrelevant, \mbox{$K_{\rho}=0.7$}, (c) a point where both the potential scattering and two particle-hole scattering processes are relevant \mbox{$K_{\rho}=0.2$}, and finally (d) when all interactions are irrelevant \mbox{$K_{\rho}=2.0$}. Here we also set \mbox{$\tilde{V}(0)/\tilde{J}(0) = 50$} initially. For the non-interacting point, Fig.~\ref{fig:7}(a), Eq.~\eqref{eq:RG_equations_final} predicts that the magnetic and potential scattering sectors are decoupled. Furthermore, the potential scattering term is marginal. Thus we do not observe any coupling between the Kondo effect and the potential scattering. In Fig.~\ref{fig:7}(b-c) the presence of a finite potential scattering leads to a strong deviation from the ideal RG flow in the absence of potential scattering. In this case there are two possibilities. In Fig.~\ref{fig:7}(b) the magnetic interactions still diverge first leading to a Kondo phase, albeit at a lower temperature. In Fig.~\ref{fig:2}, such regions where the magnetic interactions diverge first is coloured yellow. In Fig.~\ref{fig:7}(c) the presence of potential scattering leads to a divergent two-particle-hole scattering, \mbox{$V_{2,0}$}. Thus the system does not flow to a Kondo fixed point, but to a regime of strong potential scattering. This is highlighted by the purple regions in Fig.~\ref{fig:2}. In (d), on the other hand, all the interactions flow towards zero and remain perturbative. These possibility is marked by the yellow regions in Fig.~\ref{fig:2}.

\section{Poor man's scaling for a non-interacting 1D-model}
\label{app:non_interacting}

In this Appendix, we examine the reduction of the Kondo temperature from potential scattering in a non-interacting one-dimensional spin-1/2 Fermi gas using scattering theory and poor man's scaling~\cite{Anderson1970, Furusaki1994}. We consider the Hamiltonian
\begin{equation}
    H = \int dx \psi_{s}^{\dagger}(x) \left[ -\frac{1}{2}\nabla^2 + V\delta(x)\right]\psi_{s}(x) \, ,
\end{equation}
where we set the atomic mass and \mbox{$\hbar$} to unity.
The scattering states for an initial \mbox{$r$} moving fermion in this system can be readily evaluated
\begin{equation}
    \psi_{\sigma, r}(k) = \frac{1}{\sqrt{L}}\begin{cases}
        e^{i r k x} + r_ke^{-i r k x}  & rx<0 \\
        t_k e^{i r k x} & rx>0,
    \end{cases}
\end{equation}
where \mbox{$L$} is the system size, and \mbox{$r_k$} and \mbox{$t_k$} are the reflection and transmission coefficients. They are given by
\begin{align}
    |r_k|^2 &=  
    \frac{1}{1+ \frac{2k^2}{V^2}} & |t_k|^2&= \frac{1}{1+ \frac{V^2}{2 k^2}}.
\end{align}
These scattering states are the ideal basis for examining the effects of a magnetic impurity
\begin{equation}
    \delta H = J \vec{S} \cdot \psi_{s}^{\dagger}(0) \vec{\sigma}_{s,s'}^{\phantom\dagger}\psi_{s'}^{\phantom\dagger}(0) \ .
    \label{app:eq:deltaH_magnetic}
\end{equation}
Writing Eq.~\eqref{app:eq:deltaH_magnetic} in terms of the basis of scattering states gives
\begin{equation}
    \delta H = J \vec{S} \cdot \sum_{k,\ell} t_k^{\dagger}t_{\ell}^{\phantom\dagger} c_{s}^{\dagger}(k) \vec{\sigma}_{s,s'}^{\phantom\dagger}c_{s'}^{\phantom\dagger}(\ell) \ .
    \label{app:eq:deltaH_scattering}
\end{equation}
Eq.~\eqref{app:eq:deltaH_scattering} is amenable to a poor-man's scaling analysis~\cite{Anderson1970,Furusaki1994}. Restricting ourselves to scattering near the Fermi surface \mbox{$k\sim k_F$} one finds an effective antiferromagnetic coupling strength
\begin{equation}
    J_{eff} = J |t_{k_F}|^2 \ ,
\end{equation}
and a reduced Kondo temperature of the form:
\begin{equation}
    T_K = \Lambda e^{-2\pi v_F/J_{eff}} \ ,
\end{equation}
where \mbox{$\Lambda$} is the ultraviolet cutoff for the theory and \mbox{$v_F$} is the Fermi velocity. From this result, one can show that the Kondo temperature is reduced as
\begin{equation}
    \frac{T_K}{T_K^0} = e^{-\frac{\pi V^2}{Jv_F}} \approx 1- \frac{\pi V^2}{Jv_F} \ .
\end{equation}
A similar exponential supression of the Kondo temperature was obtained for the Kondo effect in the presence of fermionic systems with finite Fermi-Hubbard interactions in higher dimensions~\cite{Schork1994,Li1995,Fabrizio1995,Egger1996}. Here, we note that in the case of Fermi-Hubbard interactions, assuming that the low-temperature state is a Fermi liquid, this reduction in temperature is related to the renormalization of the density of states at the Fermi energy.

\bibliography{Luttinger_Liquid_Bib} 

\end{document}